\documentstyle[12pt,fleqn]{article}
\begin{document}

\author{{\bf Shahin S. Agaev}\thanks{e-mail address: azhep@lan.ab.az}\\
{\it High Energy Physics Lab., Baku State University,}\\
{\it Z.Khalilov st.23, 1370148 Baku, Azerbaijan}}
\title{{ \bf THE RUNNING COUPLING METHOD WITH NEXT-TO-LEADING ORDER ACCURACY AND 
PION, KAON ELM FORM FACTORS
}}
\date{}

\maketitle
\begin{abstract}
The pion and kaon electromagnetic form factors $F_M(Q^2)$ are calculated at
the leading order of pQCD using the running coupling constant method. In
calculations the leading and next-to-leading order terms in $\alpha
_S((1-x)(1-y)Q^2)$ expansion in terms of $\alpha _S(Q^2)$ are taken into
account. The resummed expression for $F_M(Q^2)$ is found. Results of numerical
calculations for the pion (asymptotic distribution amplitude) are presented.
\end{abstract}
\newpage

During last years a considerable progress were made in understanding of
infrared (ir) renormalon effects in various inclusive and exclusive
processes [1-4]. It is well known that all-order resummation of ir
renormalons corresponds to the calculation of the one-loop Feynman diagrams
with the running coupling constant $\alpha _S(-k^2)$ at the vertices, or to
calculation of the same diagrams with non-zero gluon mass. Both these
approaches are generalization of the Brodsky, Lepage and Mackenzie (BLM)
scale-setting method [5]. Studies of ir renormalon effects have also opened
interesting prospects for evalution of higher twist corrections to
processes' different characterictics [6].

As was proved in our works [2-4], exclusive processes have additional
source of ir renormalon corrections. Indeed, integration over longitudinal
fractional momenta of hadron constituents in the expression, for example, of
the electromagnetic (elm) form factor generates ir renormalon effects.

In this letter we extend our previous consideration of the pion and kaon elm
form factors in the context of the running coupling constant method.

A meson $M$ form factor in the framework of pQCD has the following form [7]

\begin{equation}
F_M\left( Q^2\right) ={\int_0^1 }
{\int_0^1 }dxdy\phi _M^{*}\left(y,Q^2\right) T_H\left(
x,y;Q^2,\alpha _S\left( {\hat Q}^2\right) \right) \phi _M\left(x,Q^2\right).\label{1} 
\end{equation}
Here $Q^2=-q^2$ is the square of the virtual photon's four-momentum, $\phi
_M\left( x,Q^2\right) $ is the meson distribution amplitude. In Eq.(1) $%
T_H\left( x,y;Q^2,\alpha _S\left( {\hat Q}^2\right) \right) $
is the hard-scattering amplitude of the subprocess $q\overline{q}'+\gamma ^{*}
\rightarrow q\overline{q}'$,
which at the leading order of pQCD is [7]

\begin{equation}
T_H\left( x,y;Q^2,\alpha _S\left( {\hat Q}^2\right) \right) = 
\frac{16\pi C_F}{Q^2}\left[ \frac 23\frac{\alpha _S\left( \left(
1-x\right) (1-y)Q^2\right) }{(1-x)(1-y)}+\frac 13 \frac{\alpha
_S(xyQ^2) }{xy}\right], \label{2}
\end{equation}
In Eq.(2) $C_F= {4\over 3}$ is the color factor, ${\hat Q}^2$ is 
taken as the square of the momentum
transfer of the exchanged hard gluon in corresponding Feynman diagrams for $%
F_M\left( Q^2\right) .$ Unlike the frozen coupling approximation, where for
calculation of $F_M\left( Q^2\right) $ one fixes the argument of $\alpha
_S\left( \left( 1-x\right) (1-y)Q^2\right) \rightarrow \alpha _S\left(
Q^2/4\right) ,$ in the running coupling method we use Eq.(2) in Eq.(1) and,
as a result, encounter with new troubles. Indeed, it is evident that $\alpha
_S\left( {\hat Q}^2\right) $ in Eq.(2) suffers from ir
singularities associated with the behaviour of the $\alpha _S\left( 
{\hat Q}^2\right) $ in the regions $x\rightarrow 0,y\rightarrow
0;$ $x\rightarrow 1,y\rightarrow 1.$ Therefore, $F_M\left( Q^2\right) $ can
be found only after proper regularization of $\alpha _S\left( {\hat Q}^2\right) $ 
in these end-point regions. For solving of this problem
let us relate the running coupling $\alpha _S\left( \lambda Q^2\right) $ in
terms of $\alpha _S\left( Q^2\right) $ by means of the renormalization group
equation. The renormalization group equation for the running coupling $%
\alpha \equiv \alpha _S/\pi $ has the form%

\begin{equation}
\frac{\partial \alpha (\lambda Q^2)}{\partial \ln \lambda }=-b_2\left[ \alpha
(\lambda Q^2)\right] ^2-b_3\left[ \alpha (\lambda Q^2)\right] ^3, \label{3} 
\end{equation}
where

$$
b_2=\frac 1{12}\left( 33-2n_f\right) ,b_3=\frac 1{48}\left( 306-38n_f\right).
$$
The solution of Eq.(3) with initial condition $\alpha (\lambda )\mid
_{\lambda =1}=\alpha \equiv \alpha _S(Q^2)/\pi $ is [8]

\begin{equation}
\frac{\alpha (\lambda )}\alpha =\left[ 1+\alpha b_2\ln \lambda -\frac{\alpha
b_3}{b_2}\left( \frac{\ln \alpha (\lambda )}\alpha -\ln \frac{b_2/b_3+\alpha
(\lambda )}{b_2/b_3+\alpha }\right) \right] ^{-1}. \label{4}
\end{equation}
This transcendental equation can be solved iteratively by keeping the
leading $ \alpha ^k\ln {}^k\lambda $ and next-to-leading $
 \alpha ^k\ln {}^{k-1}\lambda $ powers. For $\lambda
=(1-x)(1-y)$ these terms are given by%
\begin{eqnarray}
\lefteqn{\alpha \left( (1-x)(1-y)Q^2\right) \simeq \frac{\alpha (Q^2)}{1+\alpha
(Q^2)b_2\ln (1-x)(1-y)}} \nonumber \\
 & & -\frac{\alpha ^2(Q^2)b_3\ln \left[ 1+\alpha
(Q^2)b_2\ln (1-x)(1-y)\right] }{b_2\left[ 1+\alpha (Q^2)b_2\ln
(1-x)(1-y)\right] ^2}.\label{5} 
\end{eqnarray}
The first term in Eq.(5) is the solution of the renormalization group
equation (3) with leading power accuracy, whereas the whole expression (5) is
the solution of Eq.(3) with next-to-leading power accuracy.

In our previous papers [3,4] for calculation of the pion and kaon elm form
factors we used only the first term from Eq.(5). In this work for evaluation
of $F_M(Q^2)$ we use both of them. Let us clarify our approach by
considering the pion form factor and pion's simplest distribution amplitude $
\phi _{asy}(x)$
\begin{equation}
\phi _{asy}(x)=\sqrt{3}f_\pi x(1-x). \label{6}
\end{equation}
Generalization of obtained results for the pion's other distribution
amplitudes as well as for the kaon is straightforward.

After substitution of Eqs.(5),(6) into Eq.(1), the pion form factor takes
the form

\begin{eqnarray}
\lefteqn{Q^2F_\pi (Q^2)\simeq \frac{\left( 8\pi f_\pi \right) ^2}{b_2}
{\int_0^1 }{\int_0^1 }\frac{xydxdy}{t+\ln
(1-x)+\ln (1-y)}} \nonumber \\ 
    & & -\left( \frac{8\pi f_\pi }{b_2}\right) ^2\frac{b_3}{b_2}
{\int_0^1 }{\int_0^1 }\frac{xy\left\{ \ln
\left[ t+\ln (1-x)+\ln (1-y)\right] -\ln t\right\} dxdy}{\left[ t+\ln
(1-x)+\ln (1-y)\right] ^2}, \label{7}
\end{eqnarray}
where $t=1/\alpha b_2.$

The resummed expression for $F_\pi (Q^2)$ can be found using an approach
advocated in Ref.[3]. For these purposes it is instructive to change
variables from $x,y$ in Eq.(7) to $z=\ln (1-x),w=\ln (1-y)$ and apply the
inverse Laplace transforms [9]

\begin{equation}
\frac 1{\left( t+z+w\right) ^{\nu}}=\frac 1{\Gamma(\nu)}{\int_0^\infty }
\exp \left[ -u\left( t+z+w\right) \right] u^{\nu-1}du, ~Re\nu>0, \label{8}
\end{equation}
and

\begin{equation}
\frac{\ln \left( t+z+w\right) }{\left( t+z+w\right) ^2}=
{\int_0^\infty }\exp \left[ -u\left( t+z+w\right) \right] \left(
1-C-\ln u\right) udu. \label{9}
\end{equation}
In Eq.(9) $C\simeq 0.577216$ is the Euler-Mascheroni constant. After some manipulations
one gets

\begin{eqnarray}
\lefteqn{Q^2F_\pi (Q^2)\simeq \frac{\left( 8\pi f_\pi \right) ^2}{b_2}
{\int_0^\infty }\exp (-tu)B\left[ Q^2F_\pi \right] (u)du} \nonumber \\ 
 & & -\left( \frac{8\pi f_\pi }{b_2}\right) ^2\frac{b_3}{b_2}
{\int_0^\infty }\exp (-tu)u\left[ 1-C-\ln t-\ln u\right] B\left[
Q^2F_\pi \right] (u)du, \label{10}
\end{eqnarray}
where $B\left[ Q^2F_\pi \right] (u)$ is given by expression

\begin{equation}
B\left[ Q^2F_\pi \right] (u)=\frac 1{(1-u)^2}+\frac 1{(2-u)^2}-\frac
2{1-u}+\frac 2{2-u}, \label{11}
\end{equation}
and is the Borel transform of the pion elm form factor obtained in our work [3].
It is worth noting that Eq.(10) is the general expression valid for both the
pion and kaon; one needs only to replace in Eq.(10) the Borel transform $%
B\left[ Q^2F_\pi \right] (u)$ and $f_\pi $ with corresponding ones $B\left[
Q^2F_K\right] (u)$ and $f_K$ from Ref.[4]. The first term in Eq.(10) is
the pion form factor found in Ref.[3] at the leading order of pQCD in the
framework of the running coupling method using only the first term from
Eq.(5). Of course, the second term in Eq.(10) is the contribution to the form
factor coming from the second term of Eq.(5). But interpretation of the
integrand in this term (without $\exp (-tu)$) as a "traditional"
Borel transform of corresponding perturbative series is problematical,
because some regularization prescription at $u=0$ for recovering of
perturbative series is needed. Here we bypass this problem and
concentrate on calculation of the resummed expression for $F_M(Q^2)$. As we
shall see later, integration in Eq.(10) using the principal value prescription [8,10]
removes ir renormalon poles at $u=1,2...$ and gives correct results for $%
\left[ Q^2F_M(Q^2)\right] ^{res}$.

For the pion we get

\begin{eqnarray}
\lefteqn{\left[ Q^2F_\pi (Q^2)\right] ^{res}=\left[ Q^2F_\pi (Q^2)\right]
_1^{res}-\left( \frac{8\pi f_\pi }{b_2}\right) ^2\frac{b_3}{b_2}\left\{
(C+\ln t)\left[ f_{1}(0)+f_{2}(0)\right] \right.} \nonumber \\  
  & & +\left. (1-C-\ln t)\left[ (t-2)f_1(1)+(t+2)f_2(1)\right]
+g_1(0)+g_2(0) \right.\nonumber \\
  & & - \left. (t-2)g_1(1)-(t+2)g_2(1)\right\} . \label{12}
\end{eqnarray}
Here $\left[ Q^2F_\pi (Q^2)\right] _1^{res}$ is the resummed elm form factor
of the pion obtained in Ref.[3] by means of the leading term of Eq.(5),

\begin{equation}
\left[ Q^2F_\pi (Q^2)\right] _1^{res}=\frac{\left( 8\pi f_\pi \right) ^2}{b_2%
}\left[ -\frac 32+(t-2)f_1(0)+(t+2)f_2(0)\right] . \label{13}
\end{equation}
In Eqs.(12,13) we introduce following notations,
$$
f_k(n)=P.V.{\int_0^\infty }\frac{\exp (-tu)u^ndu}{%
k-u},
$$
\begin{equation}
~~~~~~~~~~~~~~~~~~~~~~~g_k(n)=P.V.{\int_0^\infty }\frac{\exp (-tu)\ln
(u)u^ndu}{k-u}. \label{14}
\end{equation}
For the pion's Chernyak-Zhitnitsky (CZ) distribution amplitude [11] in the form%
$$
\phi _{CZ} \left( x\right) =5\phi _{asy}(x)\left( 2x-1\right) ^2, 
$$
we get%
$$
\left[ Q^2F_\pi (Q^2)\right] ^{res}=\left[ Q^2F_\pi (Q^2)\right]
_1^{res}-\left( \frac{40\pi f_\pi }{b_2}\right) ^2\frac{b_3}{b_2}\left\{
\left( 1-C-\ln t\right) \left[ (t-\frac{14}3)f_1(1) \right. \right.  
$$
$$
+\left. 25(t-2)f_2(1)+8(8t+1)f_3(1)+4(4t+\frac{35}3)f_4(1)\right] +(C+\ln
t)[f_1(0)+25f_2(0) 
$$
$$
+\left. 64f_3(0)+16f_4(0)]+ g_1(0)+25g_2(0)+64g_3(0)+16g_4(0)-(t-\frac{14}%
3)g_1(1) \right.
$$
$$
\left. -25(t-2)g_2(1)-8(8t+1)g_3(1)-4(4t+\frac{35}3)g_4(1)\right\} , 
$$
with
$$
\left[ Q^2F_\pi (Q^2)\right] _1^{res}=\frac{\left( 40\pi f_\pi \right) ^2}{%
b_2}\left[ -\frac{233}6+(t-\frac{14}%
3)f_1(0)+25(t-2)f_2(0) \right.
$$
$$
\left.+ 8(8t+1)f_3(0)+4(4t+\frac{35}3)f_4(0)\right] . 
$$
Expressions for the resummed pion form factor obtained using more general
than the asymptotic and CZ distribution amplitudes [12] are rather
lengthy and will be published elsewhere.
\newpage
For the kaon we find

\begin{eqnarray*}
\lefteqn{\left[ Q^2F_K(Q^2)\right] ^{res}=\left[ Q^2F_K(Q^2)\right] _1^{res}} \\
 & & -\left(\frac{40\pi f_\pi }{b_2}\right) ^2\frac{b_3}{b_2}
\sum_{k=1}^5\left\{ \left( 1-C-\ln t\right) (t{\bf m}_k+{\bf n}_k)f_k(1)\right.  \\ 
  & & +\left. (C+\ln t){\bf m}_kf_k(0)-(t{\bf m}_k+{\bf n}_k)g_k(1)+{\bf m}_k
g_k(0)\right\} . 
\end{eqnarray*}
The values of $\left({\bf m}_k ,~{\bf n}_k\right)$ as well as $\left[ Q^2F_K(Q^2)\right]
_1^{res}$ can be found in Ref.[4].

As an example, results of numerical calculations for $\left[Q^2F_{\pi}(Q^2)\right]^{res}$
carried out using the pion's asymptotic distribution
amplitude are shown in Fig.1. Here the ratio $R=[Q^2F_{\pi}(Q^2)]^{res}/
[Q^2F_{\pi}(Q^2)]^0$, where $[Q^2F_{\pi}(Q^2)]^0$ is the pion elm form factor
calculated in the frozen coupling approximation using the same distribution
amplitude, is depicted. As is seen, the next-to-leading order correction changes
the shape of the curve in the region of small $Q^2$ $(2~ GeV^2 \leq Q^2 < 6~ 
GeV^2)$, where the whole result is smaller than the leading order one. At
$Q^2=2~ GeV^2$ the correction amounts to $\sim 15\%$ of the leading order result
decreasing toward $Q^2=6~GeV^2$. At the end of the considered values of $Q^2$
the next-to-leading order correction is positive and equals to $3-6\%$ of the 
leading order result. The similar calculations can be also fulfilled for the
pion using its alternative distribution amplitudes [12] and for the kaon. 

As was 
pointed out in Refs.[3-5],
the ir renormalon corrections can be hidden into the scale of $\alpha_S(Q^2)$ at
the leading order elm form factor $[Q^2F_M(Q^2)]^0$. In the studied case of the
pion (asymptotic distribution amplitude) we find
$$
\alpha_S(Q^2)\rightarrow \alpha_S(e^{f(Q^2)}Q^2),
$$
\begin{equation}
~~~~~~~~~~~~~~~~~~~~~~~~f(Q^2)\simeq -6.71 + 16.67\alpha_S(Q^2), \label{15}
\end{equation}
where in numerical fitting we have used Eq.(12).

It is known [3,4] that ir renormalon effects enhance the perturbative predictions
for the pion, kaon elm form factors approximately two times. Our recent studies
confirm that next-to-leading order term in Eq.(5) does not change the picture
considerably. This feature of ir renormalons may help one in solution of a contradiction
between theoretical interpretations of experimental data for the photon-to-pion
transition form factor $F_{\gamma \pi}(Q^2)$ [13] from one side and for 
the pion elm form factor $F_{\pi}(Q^2)$ [14] from another side. Thus, in works
[15,16] the authors noted that the scaling and normalization of the photon-to-pion
transition form factor tends to favor of the pion asymptotic-like distribution 
amplitude. But then prediction for $F_{\pi}(Q^2)$ obtained using the same 
distribution amplitude is lower than the data by approximately a factor of 2.
We think that in this discussion a crucial point is a chosen method of integration in Eq.(1).
Indeed, unlike $F_{\pi}(Q^2)$ the expression for $F_{\gamma \pi}(Q^2)$
at the leading order of pQCD does not contain an 
integration over $\alpha_S(xQ^2)$. In other words, the running coupling
method being applied to Eq.(1) and to the expression for $F_{\gamma \pi}$
(see, Refs.[15,16]) enhances the perturbative result for the pion elm form
factor and, at the same time, does not change $F_{\gamma \pi}$\footnote{Our statements 
concerning $F_{\gamma \pi}(Q^2)$ remain true also in the light of ir renormalon 
corrections computed 
for $F_{\gamma \pi}$ in Ref.[17], because they have another source than
ones considered in our work, namely, Feynman diagrams with bubble chains. At 
the leading order of pQCD the only source of ir renormalons is the running
coupling constant and integration
in Eq.(1) over longitudinal momenta of a meson $M$.}.
This allows us to suppose that in the 
context of pQCD the same pion distribution amplitude may explain experimental
data for both $F_{\gamma \pi}(Q^2)$ and $F_{\pi}(Q^2)$. This problem is a 
subject of separate publication.

\newpage
{\Large \bf REFERENCES}\vspace{5mm}\\
{\bf 1.} M.Neubert, Phys.Rev.D51 5924 (1995);\\
P.Ball, M.Beneke and V.M.Braun, Nucl.Phys.B452 563 (1995); Phys.Rev.D52 3929 (1995);\\
M.Beneke and V.M.Braun, Phys.Lett.B348 513 (1995);\\
C.N.Lovett-Turner and C.J.Maxwell, Nucl.Phys.B432 147 (1994); B452 188 (1995);\\
G.P.Korchemsky and G.Sterman, Nucl.Phys.B437 415 (1995);\\
B.R.Webber, Talk given at 27th ISMD97, hep-ph/9712236;\\
V.I.Zakharov, Talk presented at YKIS97, Kyoto, 97, hep-ph/9802416.\\
{\bf 2.} S.S.Agaev, Phys.Lett.B360 117 (1995); E.Phys.Lett.B369 379 (1996);\\
S.S.Agaev, Mod.Phys.Lett.A10 2009 (1995);\\
S.S.Agaev, Eur.Phys.J.C1 321 (1998).\\
{\bf 3.} S.S.Agaev, ICTP preprint IC/95/291, hep-ph/9611215.\\
{\bf 4.} S.S.Agaev, Mod.Phys.Lett.A11 957 (1996).\\
{\bf 5.} S.J.Brodsky, G.P.Lepage and P.B.Mackenzie, Phys.Rev.D28 228 (1983).\\
{\bf 6.} B.R.Webber, Phys.Lett.B339 148 (1994);\\
Yu.L.Dokshitzer and B.R.Webber, Phys.Lett.B352 451 (1995);\\
Yu.L.Dokshitzer, G.Marchesini and B.R.Webber, Nucl.Phys.B469 93 (1996);\\
M.Dasgupta, B.R.Webber, Nucl.Phys.B484 247 (1997),\\
M.Maul, E.Stein, L.Mankiewicz, M.Meyer-Hermann and A.Sch\"{a}fer, hep-ph/9710392,\\
E.Stein, M.Maul, L.Mankiewicz, A.Sch\"{a}fer, Preprint DFTT 13/97, hep-ph/9803342.\\
{\bf 7.} G.P.Lepage and S.J.Brodsky, Phys.Rev.D22 2157 (1980);\\
A.Duncan and H.Mueller, Phys.Rev.D21 1626 (1980);\\
A.V.Efremov and A.V.Radyushkin, Phys.Lett.B94 245 (1980).\\
{\bf 8.} H.Contopanagos and G.Sterman, Nucl.Phys.B419 77 (1994).\\
{\bf 9.} A.Erdelyi, Tables of integral transforms, McGraw-Hill Book Company,
New York 1954, v.1.\\
{\bf 10.} A.H.Mueller, Nucl.Phys.B250 327 (1985);\\
V.I.Zakharov, Nucl.Phys.B385 452 (1992).\\
{\bf 11.} V.L.Chernyak and A.R.Zhitnitsky, Phys.Rep.112 173 (1984).\\
{\bf 12.} G.R.Farrar, K.Huleihel and H.Zhang, Nucl.Phys.B349 655 (1991);\\
V.M.Braun and I.E.Filyanov, Z.Phys.C44 157 (1989).\\
{\bf 13.} The CLEO Collaboration, Cornell preprint CLNS 97/1477.\\
{\bf 14.} L.J. Bebek et al., Phys.Rev.D9 1229 (1974); D13 25 (1976); D17 1693 (1978).\\
{\bf 15.} A.V.Radyushkin, Acta Phys. Polon.B26 2067 (1995);\\
P.Kroll and M.Raulfs, Phys.Lett.B387 848 (1996).\\
{\bf 16.} S.J.Brodsky, C.-R.Ji, A.Pang and D.G.Robertson, Phys.Rev.D57 245 
(1998);\\
A.V.Radyushkin, Jefferson Lab. preprint, JLAB-THY-97-29, hep-ph/9707335.\\
{\bf 17.} P.Gosdzinsky and N.Kivel, NORDITA preprint 97/58, hep-ph/9707367.\\     
\newpage
{\Large \bf FIGURE CAPTION}\vspace{10mm}\\
{\bf Fig.1} The ratio $R=[Q^2F_{\pi}(Q^2)]^{res}/[Q^2F_{\pi}(Q^2)]^0$ as a 
function of $Q^2$. In calculation the pion asymptotic distribution amplitude
$\phi_{asy}(x)$ is used. The curve {\bf 1} is $R$ found using the whole result
for the resummed form factor Eq.(12), whereas the dashed curve is the ratio
$R=[Q^2F_\pi(Q^2)]_1^{res}/[Q^2F_{\pi}(Q^2)]^0$. The curve {\bf 2} corresponds
to $R$ with $[Q^2F_{\pi}(Q^2)]^0$ in the frozen coupling approximation and 
after scale-setting procedure (15). In calculations the QCD scale parameter
$\Lambda$ has been taken equal to $0.1~ GeV$.
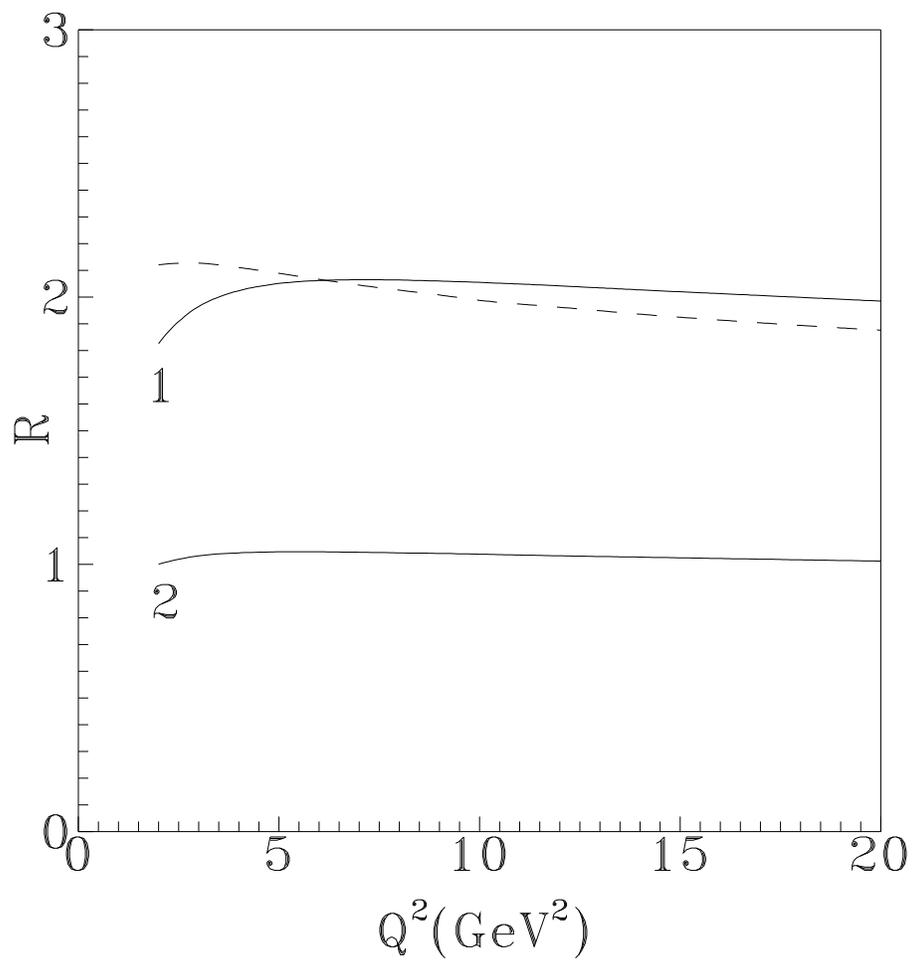
\end{document}